\documentclass[numberedappendix]{emulateapj}
\usepackage{natbib}
\shorttitle{Stability of Circumnuclear Disk in Ellipticals}
\shortauthors{Kawata, Cen, and Ho}

\begin{document}


\title{Gravitational Stability of Circumnuclear Disks in Elliptical Galaxies}

\author{Daisuke Kawata\altaffilmark{1,2},
Renyue Cen\altaffilmark{3}, and Luis C. Ho\altaffilmark{1}
}

\altaffiltext{1}{The Observatories of the Carnegie Institution of Washington,
 813 Santa Barbara Street, Pasadena, CA 91101
}
\altaffiltext{2}{Swinburne University of Technology, Hawthorn VIC 3122, Australia
}
\altaffiltext{3}{Department of Astrophysical Sciences, Princeton University, Peyton
 Hall, Ivy Lane, Princeton, NJ 08544
}

\begin{abstract}
A significant fraction of nearby elliptical galaxies are known to have 
high-density gas disks in their circumnuclear (CN) region (0.1 to a few kpc). 
Yet, ellipticals, especially luminous ones, show little signs of 
recent star formation.  To investigate the possible cause of the dearth of 
star formation in these systems, we study the gravitational stability of 
CN gas disks embedded within the gravitational potentials of both the stellar 
bulge component and the central massive black hole (BH) in elliptical galaxies.
We find that CN disks in higher mass galaxies
are generally more stable than those in lower mass galaxies,
because higher mass galaxies tend to have more massive BHs
and more centrally concentrated stellar density profiles.
We also consider the case in which 
the central stellar density profile has a core,
which is often observed for ellipticals whose total stellar mass is higher
than about $10^{11}$ $M_{\sun}$.
Such a cored stellar density profile leads to
more unstable CN disks than the power-law stellar density
profile characteristic of less massive galaxies. However, the more massive 
BHs in high-mass galaxies act to
stabilize the CN disk. Our results demonstrate that
the gravitational potentials of {\it both} the central BH and the stellar 
component should be taken into account when studying the properties of 
CN disks, as their stability is sensitive to both the BH mass
and the stellar density profile. 
Our results could explain the observed trend 
that less luminous elliptical galaxies have a greater tendency to exhibit 
ongoing star formation than giant ellipticals.
\end{abstract}

\section{Introduction}
\label{sec-intro}

 Elliptical galaxies were believed to be completely dormant and 
therefore to consist of old stellar populations, due to the galactic
wind which had blown the gas out and stopped star formation 
\citep{ja71,mb71,fg76}. The discovery
of an X-ray-emitting hot interstellar medium (ISM) in these objects has
dramatically renewed our view of their physical properties and
formation history \citep[e.g.,][]{fsjlf79,fjt85}.
Warm gas is also detected in a significant fraction of
ellipticals (55\%$-$60\%), although the estimated mass
is small ($10^{3}-10^{4}$ $M_{\sun}$) \citep{pjdsb86,hfs97a}.
In addition, multiwavelength observations 
from optical to radio reveal the existence of the fair amount of
H~{\sc I} gas \citep[e.g.,][]{ktc85,bhr92,wh94,gk99,mdzom06},
dust \citep[e.g.,][]{kgkj89,vdf95,wh95,tawti00,cmz01,ttf01,mfb04},
and molecular gas \citep[e.g.,][]{wch95,kr96,vcb03,ly05,ntksk06}
in a significant fraction of ellipticals.
Particularly, some of the cold gas appears as a disk at the center of 
the galaxies. 
High-resolution images with {\it Hubble Space
Telescope (HST)}\ uncovered that the majority of elliptical galaxies
possess such dusty cold disks, with a
typical mass of $10^{4}-10^6$ $M_{\sun}$ and sizes of
$100-500$ pc \citep[e.g.,][]{tawti00}.
Molecular gas disks are also observed in the central region,
with estimated masses of $10^{6}-10^{9}$ $M_{\sun}$ and extents up to
a few kpc \citep{gk99,okin05,ly05,ntksk06}.
In this paper we call such observed central cold structures ``circumnuclear'' 
(CN) disks. So far, there is no evidence of a correlation
between the mass of the CN disk and the total stellar mass of ellipticals
\citep[e.g.,][]{vdf95}.

 The estimated densities of the CN disks are relatively high.
CN disks in gas-rich disk galaxies obviously harbor 
star formation \citep[e.g.,][]{pjdsb86,hfs97b,kon05}. 
However, it is still a matter of debate
whether or not the CN disks in ellipticals have star formation.
\citet{hfs97a,hfs03} and \citet{lh05} suggest that 
there is no clear evidence of star formation 
in the central region of bright elliptical galaxies.
On the other hand, \citet{pjdsb86} show that 
less luminous ellipticals tend to have star formation. 
It also seems that the properties of the ISM of ellipticals 
depend on galaxy mass: the detection rates of
H~{\sc I} and molecular gas are higher in 
less luminous galaxies \citep{ls84,lkrp91,somk00,swy06}.
\citet{okin05} measured the density and rotation 
of the CN disk in the radio galaxy 3C 31, and suggest that
its disk is stable against local gravitational instability, consistent with 
the lack of evidence for star formation in this system
\citep{ook90}. \citet{ly05} also performed a similar analysis
for NGC~83 and NGC~2320. Therefore, the gravitational stability of
the disk may be a key factor for the star formation condition
in the CN disk \citep{rk89} \citep[but see also][]{kon05}.

 In the central region of elliptical galaxies, 
there are two empirical trends as a function of their mass.
One is a well-established correlation
between the black hole (BH) mass and the stellar mass of bulge
\citep{mtr98,mh03,hr04}. 
The other one is the fact that the central stellar density profile
loosely depends on the mass of galaxies
\citep{fta97,rhpfs01,smf01,fcj06,lgf07}.
The high-resolution images from the {\it HST}
enable the measurement of the stellar density profile in
the central region of galaxies 
\citep{lfh91,lfl92,lfc92,lfg93,csk93,gfl94,jfo94,lab95}.
\citet{lab95} suggest two different categories
for the inner profile of elliptical galaxies:
luminous galaxies have a ``core'' and less luminous galaxies
have a ``power-law'' profile without any evidence of a core.
On the other hand, \citet{rhpfs01} and \citet{rvdbj01} found
galaxies with ``intermediate'' inner profiles that cannot be
categorized as either cores or power laws.
\citet{teag04} attempt to link the central and global profiles of 
early-type galaxies by noting that the outer profiles of ellipticals are
well known to be fitted by the S\'{e}rsic law \citep[$r^{1/n}$ law][]{js68},
and whose index $n$ is well-correlated with the global properties
of galaxies, such as the effective radius and the total luminosity 
\citep[e.g.,][]{ccd93,gg03}.  They suggest that the inner 
power-law profile can be explained as the extension of the outer S\'{e}rsic 
profile, except for galaxies with central cores.  Therefore, among 
power-law galaxies, the brighter members have profiles with larger $n$, and 
hence steeper power-law slope for the inner profile. 
On the other hand, \citet{teag04} propose that luminous galaxies with cores 
can be described with a ``core-S{\'e}rsic'' profile,
which is a combination of a core, approximated as an inner power-law profile
with a shallow slope, and an outer
S{\'e}rsic profile.

Motivated by these facts, we study the gravitational
stability of the CN disk for elliptical galaxies with
different masses.
 The stability of a rotating disk can be
described by the so-called Toomre's $Q$-parameter \citep{at64,gl65}.
This parameter also provides a criterion for star formation 
in rotating disks \citep[e.g.,][]{rk89}.
\citet{tb05} study the stability of the CN disk analytically.
They conclude that CN disks are inevitably unstable,
and should have star formation activity even in giant ellipticals,
such as M87.  However, so far, no indication of star formation has been
observed in M87.  Although \citet{tbwb06} report the detection of the 
molecular gas in the central region of M87, the optical spectral features of 
M87's CN disk are not consistent with star formation.  The nuclear spectrum of 
M87 shows classical signatures of low-ionization nuclear emission-line regions 
\citep{hfs97c}, which as a group are consistent with being active 
galactic nuclei (AGNs) accreting at a low rate
\citep{hfs03,lh04}.

The gravitational potential in the CN region
is governed by the central BH as well as the stars in the inner bulge. 
The study of \citet{tb05}, however, ignores the potential of the stellar 
bulge component.  Our analysis of  the stability of the CN disk
in ellipticals takes into
account both contributions to the potential and 
their dependence on the global mass of the system.
Section \ref{sec-meth} summarizes our method to analyze
the stability and describes models of the CN disk
and the stellar potential in the CN region.
Section \ref{sec-res} shows the results, and in Section \ref{sec-disc}
we discuss how our results fit with the recent observed properties
of the central region of ellipticals.

\section{Method}
\label{sec-meth}

The stability of a self-gravitating disk can be
analyzed by Toomre's $Q$-parameter \citep{at64,gl65}.
The definition of Toomre's $Q$ parameter is
\begin{equation}
 Q\equiv \frac{c_{\rm s} \kappa}{\pi G \Sigma_{\rm d}},
\label{qval-eq}
\end{equation}
where $c_{\rm s}$,  $\kappa$, and $\Sigma_{\rm d}$ 
are the sound velocity, epicycle frequency, and surface density 
of the gas disk. If $Q<1$, the disk is unstable.
The specific frequency for a disk is described as
\begin{equation}
 \kappa^2 = R \frac{d \Omega^2}{d R}+4 \Omega^2,
\end{equation}
where $\Omega$ is the circular frequency.

We assume that the disk is steady, i.e. the accretion rate
is the same at different radii. The accretion rate 
can be written as
\begin{equation}
 \dot{M}(R) = -2 \pi R \alpha_{\rm acc} c_{\rm s}
 [\Omega(R) '/\Omega(R)^2] \Sigma_{\rm d}(R)
 = {\rm const.},
\label{macc-eq}
\end{equation}
 where $\alpha_{\rm acc}$ is the dimensionless viscosity parameter
\citep{ss73,jp81,fkr02}.
 This allows us to derive the density profile of the CN disk,
$\Sigma_{\rm d}(R)$, once the total mass, $M_{\rm d}$,
and the radius, $R_{\rm d}$, of the disk are fixed.
The circular density profile $\Omega(R)$ is determined by
the gravitational potential. For simplicity, we assume that the gravitational
potential is dominated by the central BH and the stellar component, and that 
the contribution from the CN disk is negligible.
As will be shown in Figure 1 below, this assumption is valid,
except for the cases of massive and compact CN disk 
in smaller mass galaxies, which are inevitably unstable.
We also assume that the stellar density
is much higher than the dark matter density in the central region.

 We examine the stability of the CN disk within the potential
of the central region of spherical galaxies with different 
total masses of the stellar bulge, $M_{\rm s}$.
We adopt the relation between the BH mass and the stellar mass 
from \citet{hr04}, 
\begin{equation}
 {\rm log} (M_{\rm BH}/M_{\sun}) = 8.20+1.12 
 {\rm log} (M_{\rm s}/10^{11} M_{\sun}).
\label{mbhmb-eq}
\end{equation}
Hence, once  $M_{\rm s}$ is fixed, we can calculate the potential
from the BH with the mass of $M_{\rm BH}$.

As mentioned in Section \ref{sec-intro}, 
the mass dependence of the inner stellar density profile is
discussed by a number of authors.
The first thorough study was carried out by the ``Nuker team,''
who introduced the ``Nuker law'' to describe the observed surface brightness
profile of the central regions of galaxies. The Nuker law \citep{lab95,fta97} is given by
\begin{equation}
 I(R)=I_{\rm b} 2^{(\beta-\gamma)/\alpha}(R/R_{\rm b})^{-\gamma}
 [1+(R/R_{\rm b})^{\alpha}]^{(\gamma-\beta)/\alpha}.
\label{nuk-eq}
\end{equation}
The asymptotic logarithmic slope inside $R_{\rm b}$ is $\gamma$; 
the asymptotic logarithmic outer slope is $\beta$; and 
$\alpha$ parametrizes the sharpness of the break. The break
radius, $R_{\rm b}$, is the point of maximum curvature
in log-log coordinates. The break surface brightness, $I_{\rm b}$,
is the surface brightness at $R_{\rm b}$.
This function is designed to fit the surface brightness profile
in the inner region, and not to describe the entire profile. 
\citet{geta03} proposed a new formula that is a combination 
of an inner power-law profile and an outer S{\'e}rsic law.
They call it the ``core-S{\'e}rsic law,'' which is described as
\begin{eqnarray}
 I(R) & = & I' \left[ 1+ (R_{\rm b}/R)^{\alpha}\right]^{\gamma/\alpha}
\nonumber
\end{eqnarray}
\begin{eqnarray}
 \exp \left\{-b\left[(R^{\alpha}+R_{\rm b}^{\alpha})/R_{\rm e}^{\alpha}
 \right]^{1/(n_{\rm s}\alpha)}\right\},
\label{cserfull-eq}
\end{eqnarray}
with
\begin{eqnarray}
 I' & = & I_{\rm b} 2^{-\gamma/\alpha}
 \exp\left[b 2^{1/(n_{\rm s}\alpha)} 
 (R_{\rm b}/R_{\rm e})^{1/n_{\rm s}}\right].
\label{cserfuli-eq}
\end{eqnarray}
The parameter $R_{\rm b}$ is the break radius
to separate the inner power law with a slope of $\gamma$
from the outer S{\'e}rsic law with effective radius $R_{\rm e}$ and
index $n_{\rm s}$, and
$I_{\rm b}$ is the surface brightness at $R_{\rm b}$.
The parameter $\alpha$ controls the sharpness of the transition
between the inner and outer profiles, where a higher value 
leads to sharper transitions.
\citet{teag04} suggest that the core-S{\'e}rsic law can be simplified
with $\alpha\rightarrow\infty$ and still provide a good description to
the observed profiles \citep[see also][]{fcj06}.  In this limit, 
\begin{eqnarray}
 I(R) & = &I_{\rm b} \left[ (R_{\rm b}/R)^{\gamma} u(R_{\rm b}-R) 
 \right. \nonumber \\
 & & \left. +e^{b (R_{\rm b}/R_{\rm e})^{1/n_{\rm s}}}
 e^{-b (R/R_{\rm e})^{1/n_{\rm s}}}
 u(R-R_{\rm b}) \right],
\label{cser-eq}
\end{eqnarray}
 where $u(x-a)$ is the Heaviside step function. 

It is still controversial which of the two formalisms better describes the 
surface brightness profile in the central region of galaxies
\citep{fcj06,lgf07}. 
In this paper, we adopt the core-S{\'e}rsic law simply for 
computational convenience to link
the density profiles in the inner region to the total mass
of the stellar bulge.
Although the simplified version of the core-S{\'e}rsic law as given in 
equation (\ref{cser-eq}) has been adopted to fit recent observations 
\citep{teag04,fcj06}, 
we use equation (\ref{cserfull-eq}) by setting $\alpha$ to 5.0,
for computational convenience.
\citet{lgf07} demonstrate how the sharp transition generated from 
equation (\ref{cser-eq}) leads to poor fits to the observed surface brightness 
profiles. The adopted $\alpha$ provides a less sharp transition.
This assumption also guarantees that the deprojected density profile does
not increase with radius with our adopted parameters (see below),
but the profiles is still close to the simplified
formula of equation (\ref{cser-eq}) used in the observations.

 As a comparison between the Nuker law (eq.~\ref{nuk-eq})
and the core-S{\'e}rsic law (eq.~\ref{cserfull-eq}),
we briefly mention how the logarithmic gradient, $\gamma'(R')$, 
at radius of $R'$ can be described for each fitting function
\citep[see also][]{teag04}.
\citet{rvdbj01} show that for the Nuker law 
\begin{equation}
 \gamma'(R')\equiv -\left[\frac{d \log I(R)}{d \log R}\right]_{R'}
 = \frac{\gamma+\beta(R'/R_{\rm b})^{\alpha}}{1+(R'/R_{\rm b})^{\alpha}}.
\end{equation}
For the core-S{\'e}rsic law (eq.~\ref{cserfull-eq}),
\citet{teag04} show that the slope can be written as
\begin{equation}
 \gamma'(R')= \frac{b}{n_{\rm s}} (1/R_{\rm e})^{1/n_{\rm s}} R'^{\alpha}
(R'^{\alpha}+R_{\rm b}^{\alpha})^{1/(n_{\rm s}\alpha)-1}
+\frac{\gamma (R_{\rm b}/R')^{\alpha}}{1+(R_{\rm b}/R')^{\alpha}}.
\end{equation}
%
%
%

 The core-S{\'e}rsic law (eq.~\ref{cserfull-eq}) is equivalent to
the S{\'e}rsic law, when $R_{\rm b}=0$ and $\gamma=0$.
Then, the S{\'e}rsic law is described as
\begin{eqnarray}
 I(x) & = & A_{\rm s} \exp(-b_{\rm s} x^{1/n_{\rm s}}),
\label{ser-eq}
\end{eqnarray}
with $x=R/R_{\rm e}$. \citet{ps97} derive the numerical solutions
\begin{eqnarray}
 A_{\rm s} & = & 
 I_{\rm tot} 
 \frac{b_{\rm s}^{2n_{\rm s}}}{2\pi n_{\rm s} \Gamma(2n_{\rm s})},\\
 b_{\rm s} & = & 2n_{\rm s}-\frac{1}{3}+0.009876/n_{\rm s},
\label{cserab-eq}
\end{eqnarray}
where $I_{\rm tot}$ is the total bulge luminosity of the
integration for $0\leq x \leq \infty$
\citep[see also][and references therein]{gd05}.
 For simplicity, we define the total bulge mass, $M_{\rm s}$,
as the total mass of the integration of the S{\'e}rsic law
of equation (\ref{ser-eq}), regardless of 
the existence of the core. 
Then, assuming a constant mass-to-luminosity
ratio $\Upsilon$ at the different radii, we can describe
the surface brightness at the break radius, $I_{\rm b}$, for 
the core-S{\'e}rsic law (eq.~\ref{cserfull-eq}) as
\begin{eqnarray}
  I_{\rm b} & =  &
  \frac{M_{\rm s}}{\Upsilon}
  \frac{b_{\rm s}^{2n_{\rm s}}}{2\pi n_{\rm s} \Gamma(2n_{\rm s})}
  e^{-b_{\rm s} (R_{\rm b}/R_{\rm e})^{1/n_{\rm s}}}.
\label{cserib-eq}
\end{eqnarray}

It is well known that $R_{\rm e}$ and $n_{\rm s}$
correlate with the total luminosity of
galaxies \citep[e.g.,][]{jk77,ccd93,ag02}.
From their fits of {\it HST}\ images
of Virgo early-type galaxies
using equation (\ref{cser-eq}), 
\citet{fcj06} 
find that the profiles of bright galaxies ($M_B\leq-20.5$ mag)
require a core, while those of less luminous galaxies 
can be described with the S{\'e}rsic law \citep[see also][]{gg03}.
They also derive the following relations between the $g$-band effective radius
and $B$-band absolute magnitude:
\begin{equation}
 \log R_{\rm e}=-0.055 (M_B+18)+1.14
\label{reser-eq}
\end{equation}
for the S{\'e}rsic galaxies, and
\begin{equation}
 \log R_{\rm e}=-0.22 (M_B+18)+1.5
\label{recser-eq}
\end{equation}
for the core-S{\'e}rsic galaxies \citep[for error estimates, see]{fcj06}.
The units of $R_{\rm e}$ is arcseconds, which for Virgo corresponds
to $1''=80.1$ pc.
In addition, they suggest that the S{\'e}rsic
index $n_{\rm s}$ and the $B$-band absolute magnitude are related by 
\begin{equation}
 \log n_{\rm s} =-0.10 (M_B+18)+0.39,
\end{equation}
valid for galaxies of both classes.
In this paper we adopt these relations for computational convenience.
Note that these relations are still not well established.
For example, \citet{gg03} and \citet{gmmdt06}
suggest a curved relation between the luminosity and
effective radius rather than two power laws.

We further assume that the stellar mass-to-light ratio for early-type galaxies 
in the $B$-band is $\Upsilon_B=M_{\rm s}/(L_B/L_{B,\sun})=7$, with 
$M_{B,\sun}=5.48$ mag, and that $\Upsilon_B$ is independent of mass; these 
assumptions are sufficiently accurate for 
early-type galaxies \citep{tbb04}. 
With some exceptions, \citet{fcj06} find that there is
a critical luminosity that separates the S{\'e}rsic galaxies from
the core-S{\'e}rsic galaxies.
With the assumptions made above, their critical luminosity of 
$M_{B,\rm c}=-20.5$ mag  
corresponds to $M_{\rm s,c}=1.73\times10^{11}$ $M_{\sun}$; 
systems with masses lower than $M_{\rm s,c}$ are assumed to follow the
S{\'e}rsic law.
Then, we derive the half-mass radius using
equations (\ref{reser-eq}) and (\ref{recser-eq}), with the assumption
that the $g$-band half-light radius is similar to the half-mass radius.
Once we fix $M_{\rm s}$, all the parameters of 
equation (\ref{cserfull-eq}) can be determined, and
equation (\ref{cserfull-eq})
provides the projected mass density profile of 
$\Sigma_{\rm s}(R)=I_B(R) \Upsilon_B$.
Assuming spherical symmetry, we can derive the three-dimensional
stellar mass density $\rho_{\rm s}(r)$ through \citep[e.g.,][]{bt87}
\begin{equation}
 \rho_{\rm s}(r) = -\frac{1}{\pi} \int^{\infty}_{r} 
 \frac{d\Sigma_{\rm s}(R)}{dR} 
 \frac{dR}{\sqrt{R^2-r^2}}.
\end{equation}
The analytical formula of three-dimensional stellar density profiles
is also discussed in \citet{ps97}, \citet{mc02},
\citet{tg05}, and references therein.

\begin{figure}
\plotone{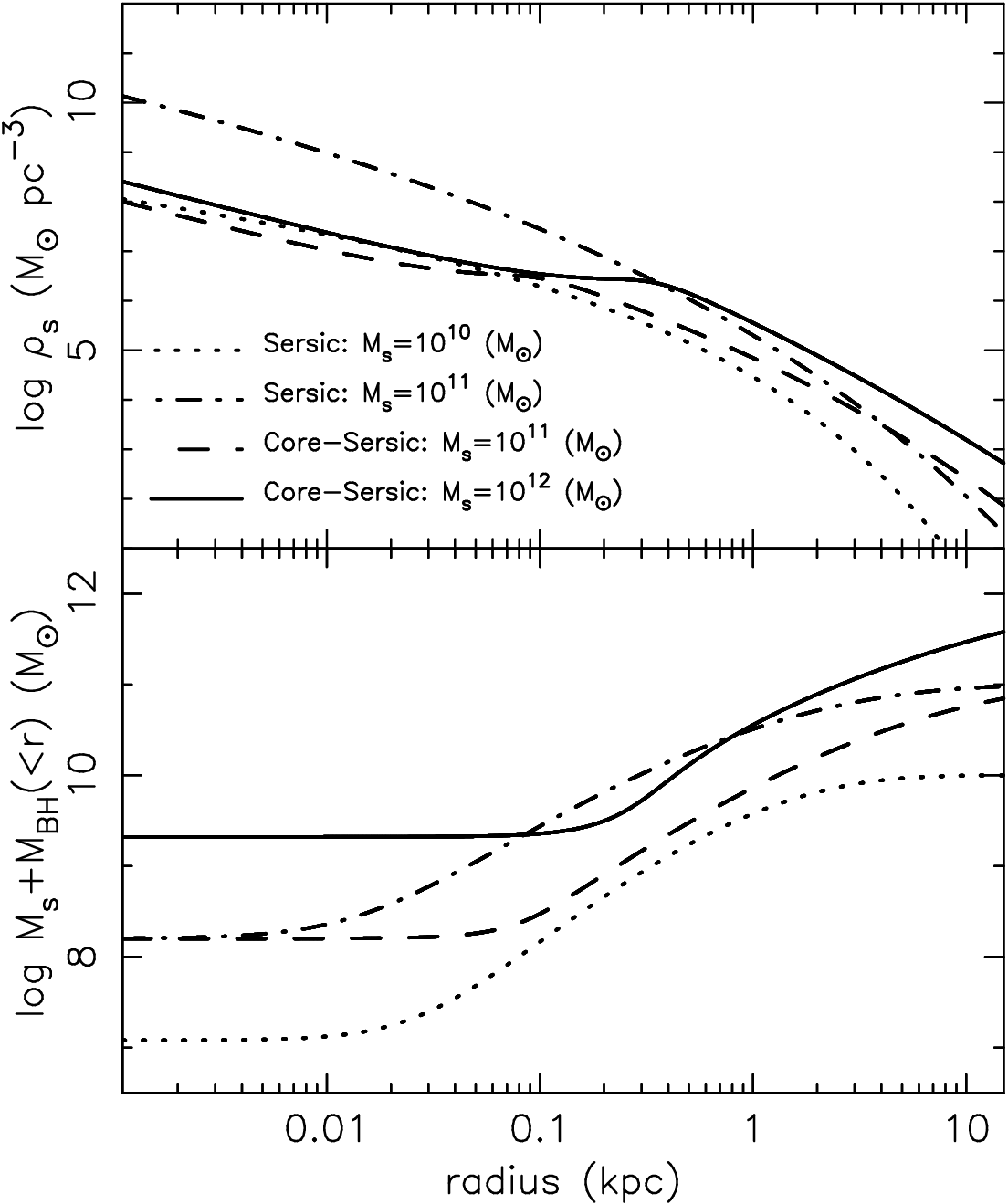}
\caption{
 Stellar density profile as a function of three-dimensional radius
({\it top}) and the total mass of stars and BH within the radius ({\it bottom})
for four different models indicated in the upper panel.
\label{fig-sprof}}
\end{figure}

\begin{deluxetable}{cccccccc}
\tablecolumns{8}
\tablewidth{0pc}
\tablecaption{Stellar Bulge Density Profiles for the
Host Galaxies \label{tab-sprof}}
\tablehead{
 \colhead{Model} &
 \colhead{$M_{\rm s}$} & \colhead{Profile} & 
 \colhead{$R_{\rm e}$} & 
 \colhead{$n_{\rm s}$} &
 \colhead{$R_{\rm b}$} & 
 \colhead{$\gamma$} &
 \colhead{$M_B$\tablenotemark{a}} \\
 \colhead{} &
 \colhead{($M_{\sun}$)} & \colhead{} &
 \colhead{(kpc)} & \colhead{} &
 \colhead{(pc)} & \colhead{} & \colhead{(mag)}
}
\startdata
S10 &
$10^{10}$ & S{\'e}rsic & 1.03 & 2.14 & $-$ & $-$ & $-17.4$
\\
S11 &
$10^{11}$ & S{\'e}rsic & 1.41 & 3.81 & $-$ & $-$ & $-19.9$
\\
cS11 &
$10^{11}$ & core-S{\'e}rsic & 6.66 & 3.81 & 9.32 & 0.1 & $-19.9$
\\
cS12 &
$10^{12}$ & core-S{\'e}rsic & 23.6 & 6.77 & 331 & 0.1 & $-22.4$
\enddata
\tablenotetext{a}{$B$-band luminosity estimated 
with a stellar mass-to-light ratio
of $\Upsilon_B=M_{\rm s}/(L_B/L_{B,\sun})=7$.}
\end{deluxetable}

\section{Results}
\label{sec-res}

\subsection{Stability of CN Disks with Fixed Mass and Radius}

As mentioned in Section \ref{sec-intro}, so far 
no correlation has been found between the mass or size of CN disks and the host galaxy mass.  In this section we study how the stability of CN disks with 
fixed mass and radius
depends on the total mass of the host galaxies.
Here, we focus on a disk with mass
${\rm log} (M_{\rm d}/{\rm M}_{\sun})=8$ 
and radius $R_{\rm d}=2.5$ kpc, which is roughly the
same as the CN disk observed in NGC~4476 by \citet{ly02}.
Disks with a range of different masses
and radii will be discussed in the next section.
Note that once $M_{\rm d}$ and $R_{\rm d}$ are fixed,
$\Sigma_{\rm d}(R)$ does not depend on $\alpha_{\rm acc}$ or $c_{\rm s}$.
Hence, $Q(R)$ described by equation (\ref{qval-eq})
is independent of $\alpha_{\rm acc}$ and is simply proportional to 
$c_{\rm s}=\sqrt{\gamma_{\rm d} k_{\rm B} T_{\rm d}/(\mu m_{\rm p})}$,
where $\gamma_{\rm d}$, $k_{\rm B}$, $T_{\rm d}$, $\mu$, and $m_{\rm p}$
are the specific heat, Boltzmann's constant, the gas disk temperature,
the mean molecular mass, and the proton mass, respectively.
Throughout the paper, we fixed $c_{\rm s}$ to a value corresponding
to $\mu=0.6$, $\gamma_{\rm d}=1$, and $T_{\rm d}=30$ K, which is
a typical temperature of observed CN disks
\citep[e.g.,][]{wch95,ttf01}.

Stability of the CN disk is examined for host galaxy models
with stellar bulge masses of 
${\rm log} (M_{\rm s}/M_{\sun})=10$, 11, and 12.
We apply the S{\'e}rsic law to the model with 
${\rm log} (M_{\rm s}/M_{\sun})=10$ (model S10),
and the core-S{\'e}rsic law to the model with 
${\rm log} (M_{\rm s}/M_{\sun})=12$ (model cS12).
Since the adopted scaling relation used in Section \ref{sec-meth} is 
known to have significant scatter and 
the model of ${\rm log} (M_{\rm s}/M_{\sun})=11$ is close to
the critical mass of $M_{\rm s,c}=1.73\times10^{11}$ $M_{\sun}$
for the two stellar density profiles, we consider two cases
of the S{\'e}rsic (model S11) law and the core-S{\'e}rsic (model cS11) 
laws for host galaxies with ${\rm log} (M_{\rm s}/M_{\sun})=11$.
All the parameter values for the stellar density
profiles and the estimated $B$-band absolute magnitude
are summarized in Table \ref{tab-sprof}.
We choose $\gamma=0.1$ for the core-S{\'e}rsic law
of equation (\ref{cserfull-eq}), which is a relatively low value
among observed cores \citep[e.g.,][]{fcj06}; this is a conservative choice 
because it leads to a maximally unstable CN disk.

 Figure \ref{fig-sprof} shows that the stellar density 
profile and the total mass of stars and BH as a function of radius
for the four models. Comparison between models S11 and S10
shows that the density profile is steeper for higher mass galaxies,
which have a larger $n_{\rm s}$.
The core-S{\'e}rsic models lead to a lower density in the inner region.
Consequently, model cS12 has even lower density than model S11 
and a similar density to model S10 within the break radius of cS12.
However, higher mass galaxies have a more massive BH.
As a result, the total mass within a given radius
is generally higher for the higher mass galaxies, 
except around the break radius for the core-S{\'e}rsic models.

\begin{figure}
\plotone{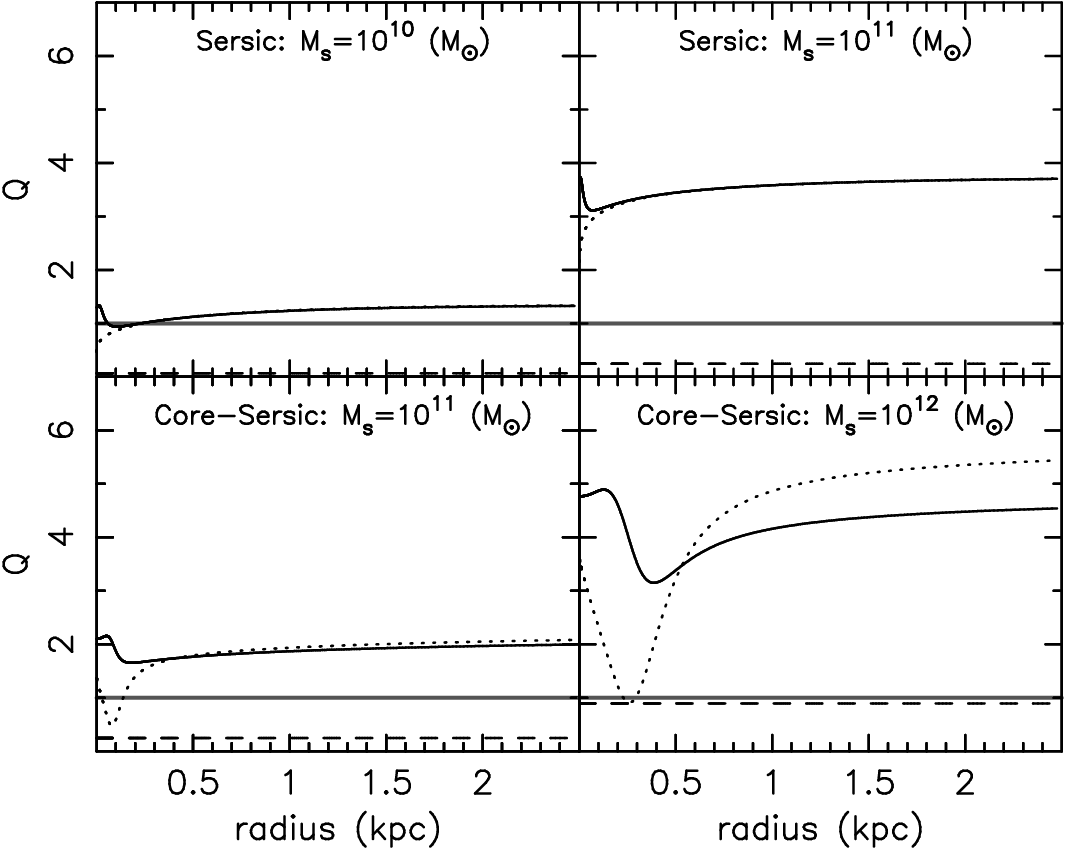}
\caption{
 Toomre's $Q$-value as a function of radius
for the CN disk with $M_{\rm d}=10^8$ $M_{\sun}$ and 
$R_{\rm d}=2.5$ kpc in galaxy models indicated in the panels. 
The black solid line presents the $Q$-value calculated by taking into 
account both the BH and the stellar potential. 
The dotted line only includes the stellar potential, while
the dashed line takes into account only the potential
of the central BH, which is correlated with
the mass of the stellar component as assumed in equation (4).
The gray solid line marks $Q=1$. 
\label{fig-qvalm8r25}}
\end{figure}

\begin{figure}
\plotone{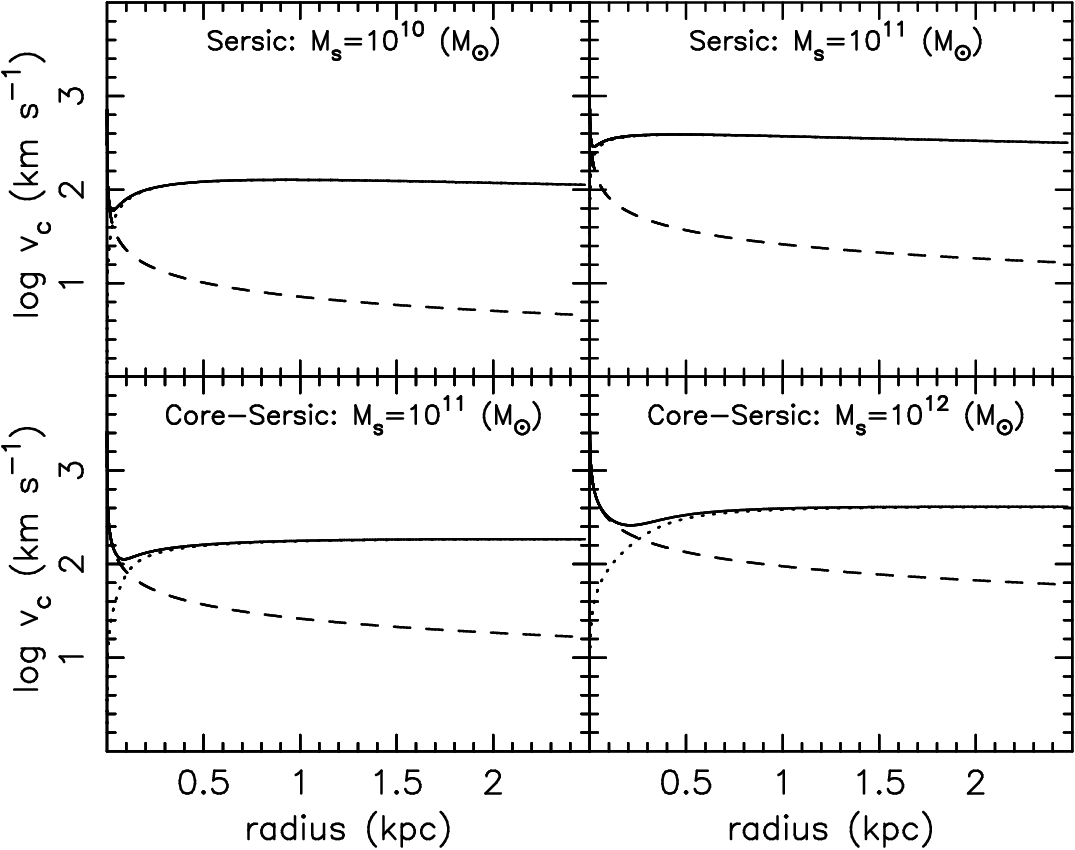}
\caption{
 The circular velocity, $v_{\rm c}$, as a function of the radius 
of the CN disk with $M_{\rm d}=10^8$ $M_{\sun}$ and 
$R_{\rm d}=2.5$ kpc in galaxy models indicated in the panels. 
The solid line takes into account both the BH and the stellar potential.  The 
dotted line only includes the stellar potential, while the dashed line takes 
into account only the potential of the central BH.
\label{fig-vcm8r25}}
\end{figure}

\begin{figure}
\plotone{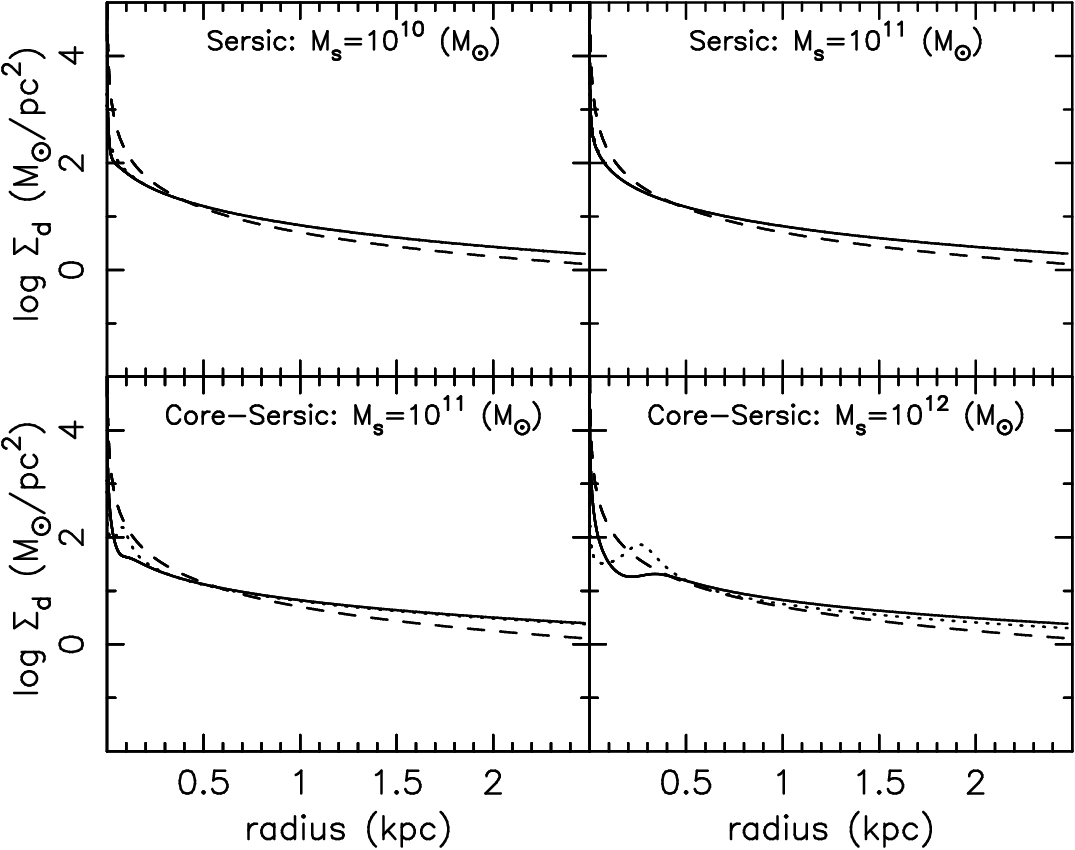}
\caption{
 The surface density profile, $\Sigma_{\rm d}(R)$, 
of the CN disk with $M_{\rm d}=10^8$ $M_{\sun}$ and 
$R_{\rm d}=2.5$ kpc in galaxy models indicated in the panels. 
The solid line takes into account both the BH and the stellar potential.  The 
dotted line only includes the stellar potential, while the dashed line takes 
into account only the potential of the central BH.
\label{fig-sigdm8r25}}
\end{figure}

 Figure \ref{fig-qvalm8r25} presents the values of Toomre's $Q$-parameter
at different radii of the CN disk for the four models. 
We also examine two additional cases for each model;
one is the case assuming no BH (dotted line), and the other one (dashed line)
ignores the stellar potential (i.e., only the Keplerian potential of the BH 
is considered).  As described in Section~\ref{sec-meth}, the 
circular velocity (Fig.~\ref{fig-vcm8r25})
is calculated purely by the BH and/or stellar potential, and the gas disk 
density profile (Fig.~\ref{fig-sigdm8r25}) is derived 
from the condition of $\dot{M}(R)={\rm constant}$ 
in equation (\ref{macc-eq}).

 The BH-only potential results (dashed lines) demonstrate that
since higher mass galaxies have a more massive BH,
CN disks are more stable in higher mass galaxies.
However, for the assumed CN disk, even model cS12
still has $Q<1$---i.e., the CN disk is unstable.
The stellar$+$BH potential results clearly demonstrate that
the stellar potential greatly helps to stabilize the CN disk. 
Moreover, comparison between the results for the stellar$+$BH potential 
and the results of the case of stellar-only potential 
indicates that the BH helps to stabilize the CN disk 
only in the inner region (a few 100 pc, depending on
the BH mass and the stellar density profiles).
Therefore, the stellar potential cannot be ignored in 
studying the dynamics of the CN disk.

For the S{\'e}rsic models, the stellar potential stabilizes 
the CN disk more in the higher mass galaxies, because higher mass
galaxies have larger $n_{\rm s}$, i.e. higher central concentration of 
the stellar potential (see also Fig.~\ref{fig-sprof}).
At a fixed stellar mass of ${\rm log} (M_{\rm s}/M_{\sun})=11$
the core-S{\'e}rsic model leads to a lower $Q$-value and a more unstable disk.
The surface density profile of the CN disk is sensitive to this 
change of the stellar potential profile (Fig.~\ref{fig-sigdm8r25}). 
The stellar-only potential results
clearly demonstrate that the disk gas density has 
a peak around the break radius.
Due to this peak density, the $Q$-value becomes 
as low as the case of the BH-only potential. However, the BH can stabilize
the disk in the inner region, and the stellar$+$BH potential case 
is much more stable.  As a result, model cS11 is more stable than model S10.

 Note that, as shown in \citet{lgf07}, the S{\'e}rsic law
often underestimates the surface brightness in the central region
where a high-density stellar component is often seen.
\citet{fcj06} invoke an additional component, which they
call "nuclei," to fit this central high-density component.
Although we do not include the nuclei in this study for simplicity,
such a compact central component also stabilizes the CN disk
in a similar way to the BH. \citet{cpf06} find that the
estimated mass of the nuclei is similar to what the BH mass and
bulge mass relation, such as equation (\ref{mbhmb-eq}), predicts. Therefore,
the effect of the nuclei would be similar to assuming 
a factor of two more massive BH.
 
 In general, CN disks in higher mass galaxies are more stabilized,
due to higher mass BH and more centrally concentrated profile of the
stellar component.
However, if the host galaxy has the core-S{\'e}rsic law profile, 
the CN disk is more unstable, compared with the CN disk 
in a galaxy with the same stellar mass and BH mass,
but having S{\'e}rsic profile. 

\begin{figure}
\plotone{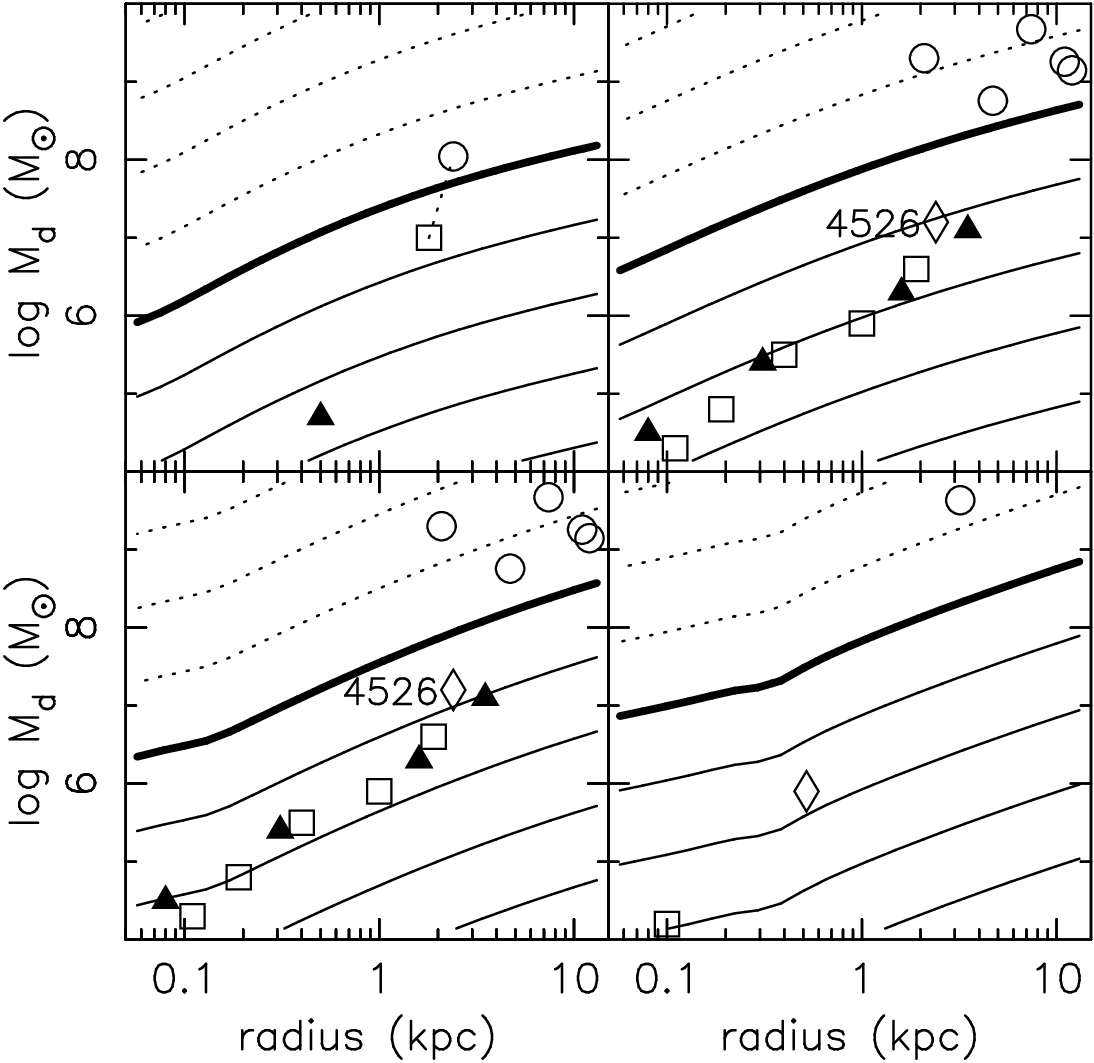}
\caption{
 The minimum $Q$-value for the CN disk with different masses and radii. The 
upper-left, upper-right, lower-left, and lower-right panels show the results 
of model S10, S11, cS11, and cS12, respectively.
The thick solid line indicates $Q=1$.
The dotted lines and thin solid lines correspond
to contours for $Q<1$ and $Q>1$.
The levels are separated by $\delta{\rm log} Q=1.0$.
Open circles denote CN disks observed in CO emission
by \citet{ly02,ly05}. The rest of the symbols show
the mass and size of the CN disk estimated from 
the optical color excess of dusty disks observed with 
{\it HST}\ \citep{tawti00}.
Solid triangles and open squares correspond to
the galaxies whose profile has been fitted with a 
S{\'e}rsic law and a core-S{\'e}rsic law, respectively 
\citep{fcj06,lgf07}; open diamonds mark galaxies with unclassified profiles.
NGC~4476, which has been observed both in CO \citep{ly02} and with {\it HST}\
\citep{tawti00}, is connected with a dotted line.
\label{fig-gqmin}}
\end{figure}

\subsection{Stability of CN Disks with Various Masses and Radii}

 In this section, we again consider the four models 
shown in Table~\ref{tab-sprof}, but study the stability of
the CN disks with different masses and radii.
Figure~\ref{fig-gqmin} shows the $Q$-value for
the CN disks with radii in the range of $0.05-15$ kpc and
masses spanning ${\rm log} (M_{\rm d}/M_{\sun})=4-10$.
Here, the figure shows the minimum $Q$-value 
in each model. For example,
$Q_{\rm min} \approx 3.15$ at $r_{\rm d}$ = 380 pc for
model cS12 in Figure~\ref{fig-qvalm8r25}.

As expected, CN disks with higher gas mass and smaller radii
are more unstable.
Comparison between S10 and S11 demonstrates that for two
CN disks with the same mass and radius in a 
S{\'e}rsic-law galaxy,
the disk in the higher mass galaxy 
is always more stable, due to the higher mass BH and 
the more centrally concentrated stellar density profile.
If the CN disk is smaller than the break radius, the highest
mass model (i.e. model cS12) is the most stable.
For a CN disk with a size comparable to the break radius, model cS12
leads to a less stable condition than model S11, 
and for the larger CN disk model cS12 is as stable as model S11.
Therefore, for large CN disks in high mass galaxies,
the size of the disk with respect to
the break radius of the host galaxies is an important factor
for the stability.


\section{Discussion and Conclusions}
\label{sec-disc}

 Our analysis of the stability of CN disks reveals
the following trends.

\begin{itemize}
\item[(1)]
 The CN disk is stabilized by the presence of the central BH.
Since higher mass galaxies tend to have more massive BHs,
the CN disk is more stable in higher mass galaxies in general.

\item[(2)]
The stellar potential is also important for the stability of
the CN disk.

\item[(3)]
 For S{\'e}rsic-law galaxies, the CN disk is more stable
in higher mass galaxies because they tend to 
have more centrally concentrated stellar density profiles.

\item[(4)]
The existence of a central stellar core in luminous ellipticals makes the CN 
disk unstable, especially around the break radius.

\end{itemize}

 As discussed in Section \ref{sec-meth}, the central surface brightness
profiles for relatively low-mass galaxies ($M_{\rm s}\leq10^{11}$ $M_{\sun}$) 
are generally described by the S{\'e}rsic law, with the trend of higher mass 
galaxies tending to have larger $n_{\rm s}$. Points (1)--(3) above 
indicate that CN disks in lower mass galaxies are more unstable. 
On the other hand, relatively high-mass galaxies
(${\rm log} (M_{\rm s}/M_{\sun})\geq11$) 
tend to have stellar density profiles that contain a central core,
which leads to a more unstable CN disk compared to the S{\'e}rsic law.
On the other hand, since such systems also have more massive BHs, 
CN disks in higher mass galaxies are kept stable, especially within the break 
radius.  These trends can explain the observational trends outlined in Section 
\ref{sec-intro}.  Because CN disks are more stable in more luminous 
ellipticals, this provides a natural explanation for central star formation to 
be curtailed in giant ellipticals, 
whereas less luminous ellipticals apparently 
have host nuclear star formation with greater ease, even though cold ISM 
in the form of dusty nuclear disks are observed to be just as prevalent in 
both environments.

 It is also worth stressing that, as seen in Figure \ref{fig-sigdm8r25},
the CN disk can remain stable even if the density of the disk is
more than 100 $M_{\sun}$ pc$^{-2}$ within 100 pc.
This is because the BH stabilizes the CN disk in the central region,
as seen in Figure \ref{fig-qvalm8r25}. 
This density is much higher than the canonical density threshold for
star formation suggested by both observations \citep{rk89} 
and theory \citep[e.g.,][]{js04},
$\Sigma_{\rm th}\sim3-10$ $M_{\sun}$ pc$^{-2}$.  Our stability analysis offers 
a simple explanation for the lack of 
star formation \citep[e.g.,][]{okin05,ly05}
in CN disks that otherwise have high densities.

 At a fixed total stellar mass for the host galaxy,
galaxies with higher $n_{\rm s}$ S{\'e}rsic law 
have more stable CN disks, while the core-S{\'e}rsic law 
leads to more unstable CN disks than the S{\'e}rsic law. 
Therefore, the frequency of central star formation 
activity should depend on both the central stellar density profile
as well as the BH mass. Recent near-ultraviolet observations
performed with {\it Galaxy Evolution Explorer (GALEX)}
indicate that some fraction of early-type galaxies 
have a small amount of recent star formation \citep{yyk05}. 
Based on semi-analytic model predictions \citet{skk06} demonstrate that the 
observed near-ultraviolet color distributions of early-type galaxies can be 
explained if there is a critical BH mass at the fixed velocity dispersion for 
galaxies that have recent star formation.
They suggested that the critical BH mass likely comes from
strong AGN heating created by the massive BH
\citep[e.g.,][]{bt95,sr98,kg05,sdh05,csw06,bbm06,co07}.
However, our study demonstrates that 
the existence of a critical condition for star formation
at the fixed velocity dispersion
can be due to the stability of the CN disk, which is 
governed by both the stellar density profile {\it and}\ the BH mass.

 As mentioned in Section \ref{sec-intro},
the size and mass of the CN disk have been measured by
several observational studies.
Some of these measurements are plotted
in Figure \ref{fig-gqmin} to compare with our model predictions.
The circles in Figure \ref{fig-gqmin} correspond to objects whose CN disk mass
and radius 
were estimated
from CO emission observed with interferometers
in \citet{ly02,ly05}.
Figure \ref{fig-gqmin} also contains objects whose disk properties were 
estimated from the color excess of dust features measured in optical images of 
the central region of early-type galaxies observed with the {\it HST}
\citep{tawti00}\footnote{\citet{tawti00} find 
that some galaxies in their sample have irregular dust morphology.
Figure \ref{fig-gqmin} only shows data for galaxies
whose dust morphology is classified a disk.}.
The stellar density profiles of the central region for
the majority of the galaxies in \citet{tawti00} have been studied
by \citet{fcj06} and/or \citet{lgf07}. This allows us to further distinguish 
the objects by their central profile type.

Following Section \ref{sec-meth}, we estimated the stellar mass from the 
$B$-band luminosity, assuming a stellar mass-to-light ratio 
of $\Upsilon_B=M_{\rm s}/(L_B/L_{B,\sun})=7$.
The $B$-band luminosities come from the NASA/IPAC Extragalactic 
Database (NED). We plot the data for galaxies with masses
${\rm log}(M_{\rm s}/M_{\sun})<10.5$ and ${\rm log}(M_{\rm s}/M_{\sun})>11.5$
in the upper-left and lower-right panels of the figure, respectively;
the rest of galaxies are shown in both the upper-right and lower-left panels.

Only one object---NGC~4476--- has been measured with both techniques 
(CO from \citet{ly02} and dust from \citet{tawti00}); the two data points are 
connected with the dotted line.  \citet{ly02} obtained a much higher gas mass 
of $1.1\times10^8$ $M_{\sun}$ within 2.4 kpc, compared with a mass of 
$1.0\times10^7$ $M_{\sun}$ within 1.8 kpc in \citet{tawti00}.  The 
order-of-magnitude discrepancy in mass cannot be explained by the slightly 
different distances adopted by these authors (18 Mpc by \citet{ly02} and 
16.8 Mpc by \citet{tawti00}).  It is noteworthy that {\it all} of the 
CO-measured CN disks have systematically larger disk masses than the 
dust-measured systems, suggesting that at least part of the discrepancy 
may be due to systematic errors in the estimated gas masses.  
Both sets of observations require a conversion factor to arrive at a gas mass.
In the case of the CO observations, a standard Galactic CO-to-H$_2$ conversion 
factor was adopted.  The CO-to-H$_2$ conversion factor, however, may be 
systematically lower in regions of high metallicity
\citep{ast96}, as is the case in the central regions of massive galaxies, 
a suggestion supported by radiative transfer calculations in numerical 
simulations of CN disks \citep{wt05}.  If this is the case, then the gas masses
from \citet{ly02,ly05} are overestimated.

A similar caveat applies to the gas masses estimated from the dust extinction, 
which assume a Galactic gas-to-dust mass ratio \citep{tawti00}.  However, 
dust mass estimated from optical extinction should be considered as lower 
limit to the true dust masses \citep{gdj95,ttf01,mfb04}.  For example, 
\citet{gdj95} show that the dust masses derived from the far-infrared emission 
are roughly an order of magnitude higher than those estimated from optical 
extinction alone.  They argue that the optical extinction may underestimate 
the dust mass by about factor of 2 due to the assumption that the dust is in 
front of the stars \citep[see also][who claim that this effect is more 
significant]{mtsb00}.  \citet{gdj95} also discuss that the discrepancy in the 
estimated dust mass is mainly due to the presence of diffusely distributed 
dust, which cannot be seen as optical extinction.  We note, however, that such 
a diffuse component is unlikely to be present as a rotating disk, as it is 
more likely to be supported by velocity dispersion.  In any case, it seems 
plausible that the gas masses derived from the optical dust features may be 
an underestimate of the true gas mass, thereby narrowing the disagreement
with the CO-based masses.  Until more accurate gas masses are available, it is
difficult to draw more quantitative comparisons between our model predictions 
and observations.

To see if there is any sign of ongoing star formation in the sample of
galaxies shown in Figure 5, we have done a careful search of the literature
to inspect published optical spectra
\citep{pjdsb86,bba89,hfs97c,crdz00,dd03}.
Galaxies with ongoing star formation show optical emission-line ratios that
are readily distinguishable from other sources of ionization 
\citep[e.g., active galactic nuclei; see][]{hfs97c}.
 Among all the objects, only NGC 4526 has a spectral classification 
consistent with stellar photoionization \citep{hfs97c}.
The rest either have no star formation (14 galaxies) or have
insufficient spectral information to tell (5 galaxies). 
Although NGC 4526, which is highlighted in Figure \ref{fig-gqmin}, 
is in the stable ($Q>1$) region, interestingly, NGC 4526 has one of
the most unstable CN disks among the {\it HST}-measured sample.
Also, none of the CO-measured CN disks, which are located
in the unstable region in Figure \ref{fig-gqmin}, show a clear indication of
star formation. Although this contradicts with our prediction,
because of the above-mentioned ambiguities in the estimates of the gas masses
from the observations, it is difficult to conclude if these galaxies
require additional process to explain their suppression of star formation.

It is clear from inspection of Figure~\ref{fig-gqmin} that once the mass
of the CN disk becomes high enough, the disk will inevitably become 
unstable for star formation, one of the consequences of which may be 
to aid gas fueling to the AGN \citep{kw04}.  This type of situation is likely 
to be realized in the aftermath of a gas-rich major merger, whereby the large 
amount of the gas dissipated toward the center can 
generate a high-mass CN disk.  High-resolution millimeter observations of 
luminous infrared galaxies indeed suggest that systems with 
larger central gas surface densities tend to either form stars with greater
efficiency or have a higher probability of hosting an AGN \citep{bs99}.

Finally, we speculate that star formation induced in an unstable CN disk
may be directly related to the formation of kinematically decoupled cores 
(KDCs) that are often seen in the centers of early-type galaxies 
\citep[e.g.,][]{jk84,rb88,bs92,cmp00} as well as the central stellar
disk that many power-law ellipticals seem to have
\citep{lfg05}.
One of the intriguing clues regarding the origin of KDCs
is that large cores are always old \citep{mes06}.  
If KDCs are a by-product of
unstable CN disks, then large KDCs require large, massive CN disks, which 
probably can only be formed through major, gas-rich mergers, which are 
most prevalent at earlier epochs.

\acknowledgments

We thank Alister Graham, Tod Lauer, and Lisa Young for helpful comments.
This research has made use of the NASA/IPAC Extragalactic Database (NED) 
which is operated by the Jet Propulsion Laboratory, California Institute of 
Technology, under contract with the National Aeronautics and Space 
Administration. This work is supported in part by grants NNG05GK10G
and AST-0507521.
The work of L. C. H.  is supported by the Carnegie 
Institution of Washington and by NASA grants from the Space Telescope Science 
Institute (operated by AURA, Inc., under NASA contract NAS5-26555). 




\end{document}